# Status of the Milagro Gamma Ray Observatory

J.F. McCullough[1] for the Milagro Collaboration
[1]*Department of Physics, University of California, Santa Cruz, CA 95056, USA*

## Abstract

The Milagro Gamma Ray Observatory is the world's first large-area water Cherenkov detector capable of continuously monitoring the sky at TeV energies. Located in the mountains of northern New Mexico, Milagro will perform an all sky survey of the Northern Hemisphere at energies between $\sim 250\,\text{GeV}$ and 50 TeV. With a high duty-cycle ($\sim 100\,\%$), large detector area ($\sim 5000\,\text{m}^2$), and wide field-of-view ($\sim 1$ sr), Milagro is uniquely capable of searching for transient and DC sources of high-energy $\gamma$-ray emission. Milagro has been operating since February, 1999. The current status of the Milagro Observatory and initial results will be discussed.

## 1 Introduction

Observations in high-energy $\gamma$-ray astronomy can be performed with either satellite or ground-based detectors. Satellite-based telescopes directly detect photons by converting them and then tracking the electron-positron pairs. Ground-based telescopes detect the secondary charged particles in the extensive air shower (EAS) that results when an incoming photon interacts with the earth's atmosphere. Because of the low fluxes involved in high-energy $\gamma$-ray astronomy and the relatively small detectors that can be placed on satellites, observations above a few 10s of GeV must be performed from the ground.

Atmospheric Cherenkov telescopes (ACTs) have been used with great success in the energy region from $\sim 300$ GeV - 10 TeV. ACTs detect the Cherenkov radiation produced in the atmosphere from the relativistic charged secondaries in an EAS. The advantages of ACTs over other ground-based techniques is that they have a low energy threshold, very good angular resolution and excellent background rejection capabilities. However, ACTs are pointed telescopes with a small field of view and can therefore only observe one source at a time. In addition, because they are optical instruments, ACTs have a very small duty factor ($\sim 10\%$) since they can only be used on clear, dark nights.

In the energy region above $\sim 40$ TeV, enough secondary particles from an EAS reach the ground that an extensive air-shower particle detector array can be used. This typically consists of a sparse array of scintillation counters that detect the charged particles from an air shower that reach ground level. EAS arrays can observe the entire overhead sky at once and can therefore observe all sources within their field of view simultaneously. They can also be operated 24 hours a day in all weather conditions. However, EAS arrays typically cover only $< 1\,\%$ of the ground with detectors and therefore only detect a small fraction of the charged particles reaching the earth's surface. Because of this, EAS arrays have a much higher energy threshold than ACTs and have very limited background rejection capabilities.

Ideally, one would like to have the high duty factor and large aperture of EAS arrays in the energy region covered by ACTs. This would allow the first all-sky survey to be done at TeV energies. In order to accomplish this with an EAS array, one could move to a higher altitude, detect a larger fraction of the charged particles reaching ground level, or increase the sensitivity to the photons from the EAS that reach the ground. Milagro has incorporated the last two ideas to achieve an energy threshold of $\sim 250$ GeV while maintaining a high duty factor and large aperture.

## 2 Detector Design

Milagro is the first large-area water-Cherenkov detector specifically built to study extensive air showers. The detector is located in the mountains of northern New Mexico at an altitude of 2650m. Milagro is built in a man-made pond formerly used as part of a geothermal energy project. The pond is $60 \times 80\,\text{m}^2$ at the surface and has sloping sides that lead to a $30 \times 50\,\text{m}^2$ bottom at a depth of 8 m. It is filled with 5 million gallons of purified water and is covered by a light-tight high-density polypropylene liner. Milagro consists of two layers of upward pointing 8" diameter hemispherical Hamamatsu 10-stage photomultiplier tubes (PMTs). Each PMT is lifted by its buoyant force ($\sim 8$ pounds each) and is anchored by Kevlar strings to a $3 \times 3\,\text{m}^2$ support grid of 3" PVC pipe filled with wet sand. The top (air-shower) layer of 450 PMTs is located 1.4 m below the water's surface. This layer is used to trigger the detector and measure the arrival time of the air-shower wave front. The second (hadron/muon) layer consists of 273 PMTs located at a depth of approximately 7 m. The hadron layer is used to make a calorimetric measurement of the shower, to differentiate $\gamma$-induced air showers from cosmic-ray induced showers and to detect muons.

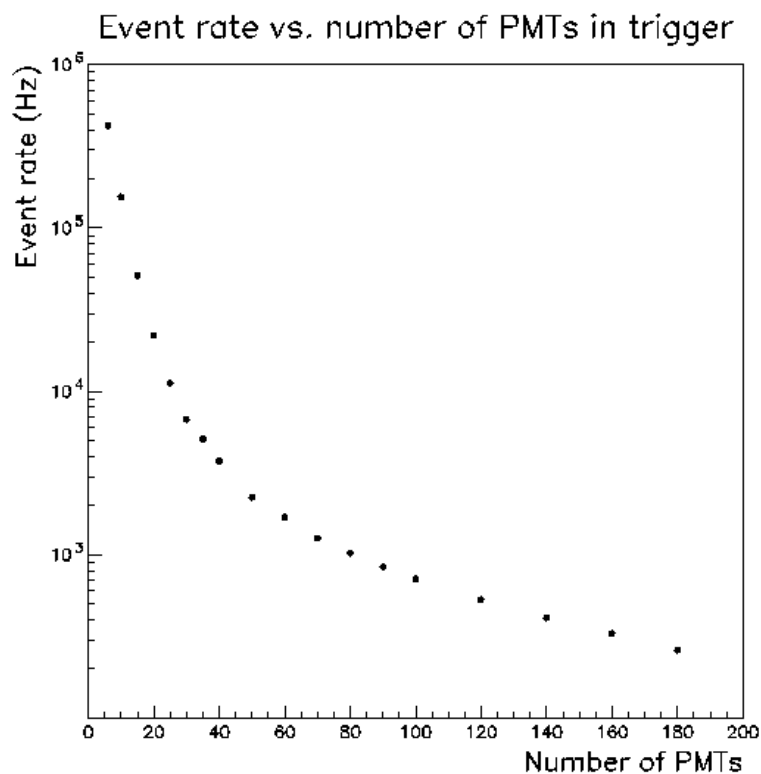

Figure 1: Event rate vs. number of PMTs required to trigger the detector

The use of water as a detection medium has several distinct advantages over EAS arrays that employ scintillation counters. At the earth's surface, there are 4-5 times more photons in an extensive air shower than charged particles. When these photons enter the water, they convert to electron-positron pairs or Compton scatter electrons; these products are subsequently detected by the Cherenkov radiation that they emit. Since the Cherenkov light cone in water is large ($\sim 42^\circ$), the radiation spreads out so that a sparse array of PMTs provides complete coverage of the entire pond. Milagro therefore provides nearly 100% coverage of the surface as compared to $< 1\,\%$ for a scintillation array. The increased sensitivity to photons and the detection of a greater fraction of the charged particles in an EAS allows Milagro to achieve a substantially lower energy threshold than scintillation arrays.

## 3 Event Reconstruction

The trigger condition currently used is a simple multiplicity of PMT hits within a coincidence window of approximately 200 ns. Figure 1. shows the event rate for Milagro as a function of the number of air-shower PMTs required to trigger the detector. For each event, the arrival time and pulse height (number of photoelectrons or PEs) for each PMT hit are recorded. From this information, a number of quantities including the direction of the incident primary, the location of the shower core, and the energy of the primary particle are reconstructed. Of these quantities, the direction of the primary is the most important since the detection of a $\gamma$-ray source is based primarily upon the observation of an excess of events above the isotropic background of cosmic-ray induced air showers from a particular region of the sky.

To determine the direction of the primary $\gamma$-ray (or cosmic-ray), Milagro employs the same technique used by conventional scintillation-counter arrays. After the primary $\gamma$-ray or cosmic-ray interacts in the atmosphere and creates an air shower, the secondary particles are all highly relativistic and therefore beamed forward in the direction of the primary. The end result (to a first approximation) is a flat pancake, perpendicular to the incident $\gamma$-ray or cosmic-ray, composed of many thousands of photons, electrons, positrons, and hadrons traveling parallel to the direction of the primary particle. By measuring the relative times that PMTs in the air-shower layer are struck by the Cherenkov radiation, the direction of the primary particle is reconstructed. An example of a reconstructed shower in Milagro is shown in Figure 2. The orientation of the fitted plane is determined by a least-squares ($\chi^2$) fit to a more complex shower-front shape using the measured times and positions of the air-shower PMTs. The angular resolution of Milagro depends upon the number of PMTs used in the fit. Monte Carlo simulations of the detector response suggest a typical angular resolution of less than $1^\circ$.

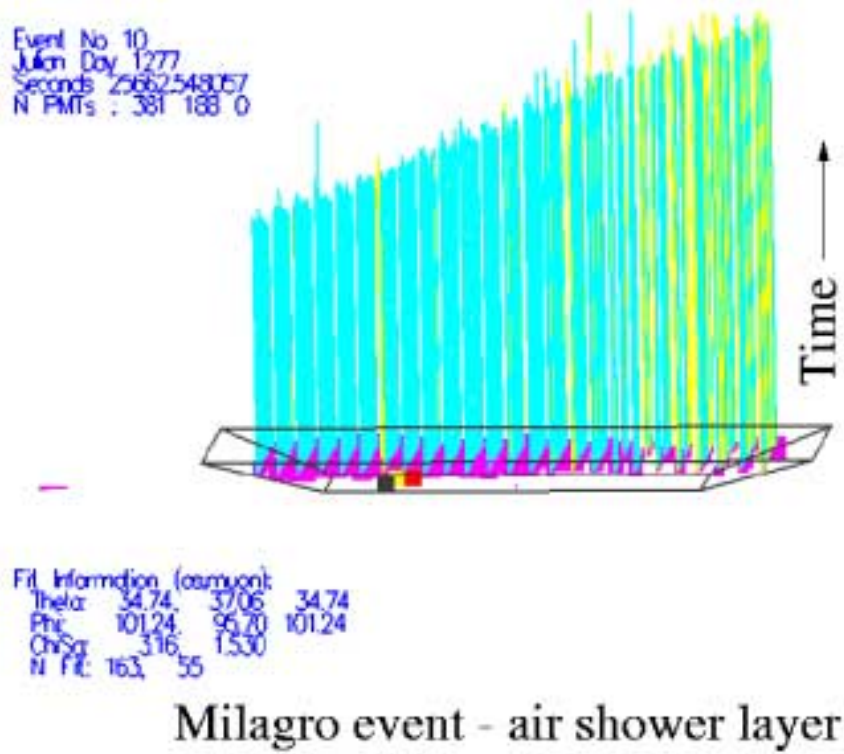

Figure 2: Event display for Milagro. Vertical lines are proportional to the arrival time

The location of the shower core and the energy of the primary particle are reconstructed from the amplitudes and distributions of pulse heights of hit PMTs in both the air shower and hadron layers. The ability to reconstruct the energy of the primary depends heavily on the ability to find the shower core. This is because a high energy shower hitting far from the pond and a low energy shower hitting close to the pond can both appear the same to Milagro, which only has information on the particles entering the water. To allow a better determination of the core location for showers which land outside the pond, a sparse array of water tanks is being deployed around Milagro. Each water tank is equipped with a PMT that detects most of the shower particles entering it. Monte Carlo simulations predict that with an array of 172 water tanks, Milagro will be able to find the shower core to within approximately 15 meters (Shoup et al., 1999).

Background rejection is accomplished using the pulse heights of the hadron layer PMTs. Muons penetrating to the hadron/muon layer leave a very distinct signal as can be seen in Figure 3. One or two PMTs are usually hit with amplitudes $\geq 20$ photo-electrons while the neighboring tubes have much lower amplitudes. Since muons are mainly produced in cosmic-ray induced air showers, any event identified as containing a muon is thrown out. We thus have an effective method of background rejection. One disadvantage with this method is that it only works if a muon strikes the pond. According to Monte Carlo simulations, this only happens in approximately $50\%$ of the proton showers which trigger Milagro. Other algorithms for identifying background events based on the distribution of light in the hadron/muon layer are promising and are currently being investigated.

## 4   Milagro Operation and Results

A prototype detector (Milagrito) was operated from February 1997 to May 1998. Milagrito was approximately half the size of Milagro ($\sim 2500\,\mathrm{m}^2$) and consisted of a single layer of 228 PMTs. Data was taken with the PMTs at depths of 1.0m, 1.5m, and 2.0m to determine the effect of water depth on angular resolution. Because Milagrito had only one shallow layer of PMTs, it had very limited background rejection. The angular resolution for Milagrito was $\sim 1^o$. Milagrito used a trigger condition of 100 PMTs hit within a coincidence window of 300ns. This resulted in an event rate of 300-400 Hz, depending on the water depth. In the 15 months of operation of Milagrito, we collected $\sim 8.9 \times 10^9$ events and wrote $\sim 9$ Terabytes of data to tape.

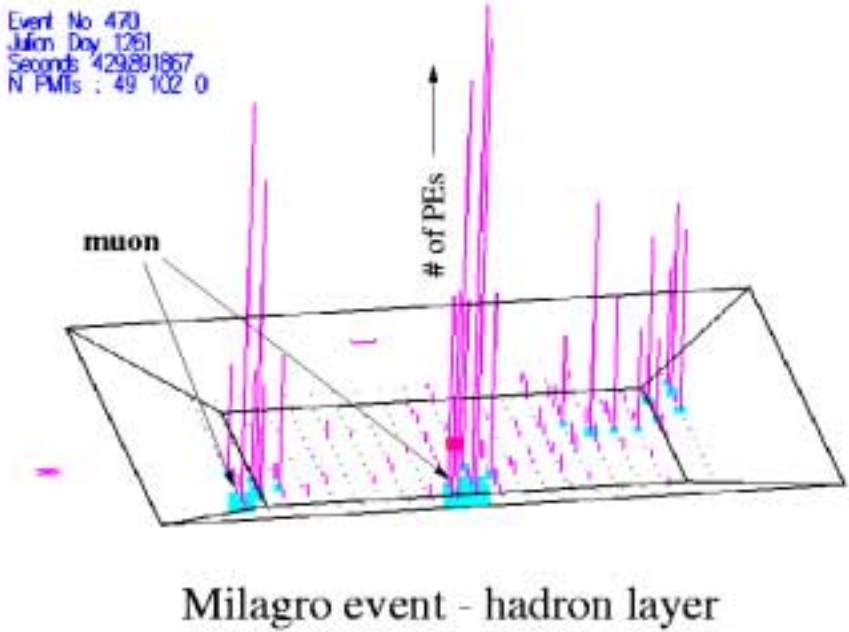

Figure 3: Event display for Milagro. Vertical lines are proportional to pulse height

Although Milagrito was operated mainly as a test run for this relatively new water-Cherenkov technique, it was a fully operational detector that has produced several interesting scientific results. Milagrito detected the moon shadow with a significance of $10\,\sigma$ (Wascko et al., 1999), detected Markarian 501 with a significance of $> 3\,\sigma$ (Westerhoff et al., 1999), and detected the Nov. 6, 1997 solar coronal mass ejection (Ryan et al., 1999). We are continuing to analyze the Milagrito data.

Milagro was installed in the summer of 1998 and began taking data in February 1999. The electronics for Milagro use the same time-over-threshhold technique used in Milagrito (Atkins et al., 1999). As of this writing, Milagro is in an engineering mode. The PMTs are being calibrated and final adjustments to the data acquisition system are being made. The Milagro trigger is currently 150 PMTs hit within 200ns. This results in an event rate of $\sim 350$ events per second and $\sim 75$ Gigabytes of data written to tape each day. We have collected $\sim 2$ billion events to date. We expect to begin normal operations in early June. Preliminary results from the Milagro data will be presented at the conference.

## Acknowledgements

This research is supported in part by the National Science Foundation, the U.S. Department of Energy Office of High Energy Physics, the U.S. Department of Energy Office of Nuclear Physics, Los Alamos National Laboratory, the University of California, the Institute of Geophysics and Planetary Physics, The Research Corporation, and the California Space Institute.

## References

Atkins et al. 1999, Nucl. Instr. Meth. A , in preparation
Ryan, J. et al. 1999, Proc. 26th ICRC (Salt Lake City, 1999) SH 1.7.02.
Shoup, A. et al. 1999, Proc. 26th ICRC (Salt Lake City, 1999) OG 4.4.06.
Wascko, M.O. et al. 1999, Proc. 26th ICRC (Salt Lake City, 1999) SH 3.2.39.
Westerhoff, S. et al. 1999, Proc. 26th ICRC (Salt Lake City, 1999), OG 2.1.11.